\documentclass[twocolumn]{aastex631}

\usepackage{subfigure}

\begin{document}

\title{Understanding the Formation and Evolution of Dark Galaxies in a Simulated Universe}

\correspondingauthor{Ho Seong Hwang}
\email{hhwang@astro.snu.ac.kr}

\author[0009-0007-8443-3143]{Gain Lee}
\affiliation{Astronomy Program, Department of Physics and Astronomy, Seoul National University, 1 Gwanak-ro, Gwanak-gu, Seoul 08826, Republic of Korea}

\author{Ho Seong Hwang}
\affiliation{Astronomy Program, Department of Physics and Astronomy, Seoul National University, 1 Gwanak-ro, Gwanak-gu, Seoul 08826, Republic of Korea}
\affiliation{SNU Astronomy Research Center, Seoul National University, 1 Gwanak-ro, Gwanak-gu, Seoul 08826, Republic of Korea}

\author[0000-0002-6810-1778]{Jaehyun Lee}
\affiliation{Korea Astronomy and Space Science Institute, 776 Daedeokdae-ro, Yuseong-gu, Daejeon 34055, Republic of Korea}

\author{Jihye Shin}
\affiliation{Korea Astronomy and Space Science Institute, 776 Daedeokdae-ro, Yuseong-gu, Daejeon 34055, Republic of Korea}

\author{Hyunmi Song}
\affiliation{Department of Astronomy and Space Science, Chungnam National University, Daejeon 34134, Republic of Korea}




\begin{abstract}

We study the formation and evolution of dark galaxies using the IllustrisTNG cosmological hydrodynamical simulation. We first identify dark galaxies with stellar-to-total mass ratios, $M_* / M_{\text{tot}}$, smaller than $10^{-4}$, which differ from luminous galaxies with $M_* / M_{\text{tot}} \geq 10^{-4}$. We then select the galaxies with dark matter halo mass of $\sim 10^9 \, h^{-1}$$\rm M_{\odot}$ for mass completeness, and compare their physical properties with those of luminous galaxies. We find that at the present epoch ($z=0$), dark galaxies are predominantly located in void regions without star-forming gas. We also find that dark galaxies tend to have larger sizes and higher spin parameters than luminous galaxies. In the early universe, dark and luminous galaxies show small differences in the distributions of spin and local environment estimates, and the difference between the two samples becomes more significant as they evolve. Our results suggest that dark galaxies tend to be initially formed in less dense regions, and could not form stars because of heating from cosmic reionization and of few interactions and mergers with other systems containing stars unlike luminous galaxies. This study based on numerical simulations can provide important hints for validating dark galaxy candidates in observations and for constraining galaxy formation models.

\end{abstract}



\section{Introduction} \label{sec:intro}

Dark galaxies are intriguing objects that are believed to be made up almost entirely of dark matter (DM), with little or no stars. The standard $\Lambda$ Cold Dark Matter ($\Lambda$CDM) cosmological model posits that galaxies are assembled by hierarchical merging of smaller dark matter halos \citep{White1978}. Galaxies are expected to form stars in a gravitational potential shaped by dark matter halos. However, some dark matter halos may fail to form stars from their gas reservoirs, only harboring unseen, resulting in the formation of dark galaxies at their center.

The study of dark galaxies is important for several reasons. Firstly, they offer a unique laboratory to test the $\Lambda$CDM model in the context of galaxy formation and evolution. Secondly, dark galaxies may provide a partial solution to the missing satellite problem, which refers to a mismatch between the predicted number of low-mass dark matter halos from the simulation and the observed number of satellite galaxies in the Local Group \citep{Klypin1999, Moore1999, Verde2002, Park2018}. Lastly, the investigation of dark galaxies could shed light on the nature of dark matter itself.

Observationally, detecting dark galaxies is challenging, especially in optical wavelength, due to their absence of stars. However, if they contain baryonic content like a gas reservoir, they could be detectable through 21-cm radiation from neutral hydrogen (HI) gas, which would make them appear as HI sources without an optical counterpart. Several candidates for dark galaxies have been proposed, including VIRGOHI21 \citep{Minchin2005}, Dragonfly 44 \citep{vanDokkum2016}, AGESVC1 282 \citep{Bilek2020}, AGC 229101 \citep{Leisman2021}, and FAST J0139+4328 \citep{Xu2023}. However, their existence remains under debate because other possible explanations, not of cosmological origin, have also been suggested. These include intergalactic gas clouds resulting from ram-pressure stripping \citep{Oosterloo2005, Junais2021} and tidal debris stemming from galaxy-galaxy interactions \citep{Bekki2005, Duc2008, Dansac2010, Taylor2022}. Therefore, further studies on the nature of dark galaxies are necessary to examine whether these candidates are genuine dark galaxies or not.

On the other hand, there have been few studies on dark galaxies from the theoretical side. One of the earliest investigations on this topic was conducted by \citet{Jimenez1997}, who suggested that dark galaxies have initially high spin parameters, resulting in larger sizes, lower surface densities, and leading to lower star formation rates. This idea was revisited by \citet{Jimenez2020}, who demonstrated that observed dark galaxy candidates are consistent with the previous prediction. However, their study was based solely on the Toomre stability criterion, a simple condition for the gravitational stability of a disk, and thus did not account for other factors, such as localized star formation activity or baryonic feedback mechanisms that could significantly impact star formation rates.

In a numerical study of \citet{Benitez2017}, they analyzed the properties of dark galaxies, represented by star-free low-mass dark matter halos in the Local Group, using the APOSTILE zoom-in simulations. They found little overlap in properties between simulated dark galaxies and observed Ultra-Compact High-Velocity Clouds (UCHVCs), which are another potential candidate for dark galaxies. This finding suggests that the large number of UCHVCs may not belong to the population of abundant dark galaxies. However, this study focuses solely on the Local Group-like environment, limiting the generalizability of the findings.

In this context, we aim to explore the nature of dark galaxies using the state-of-the-art cosmological simulation, IllustrisTNG \citep{Nelson2018}. Our research pursues two main goals: (1) to identify dark galaxies and compare their properties with those of luminous galaxies, and (2) to understand the origin and evolution of dark galaxies. Furthermore, our research expands upon previous studies by investigating dark galaxies in a broader range of environments including both clusters and fields, and by considering the impact of various baryonic feedback mechanisms on their formation.

This paper is organized as follows. In Section 2, we provide a brief overview of the IllustrisTNG simulation and clarify the selection criteria for our galaxy sample. In Section 3, we present the physical properties of dark galaxies and compare them to those of luminous galaxies in terms of gas properties, internal properties, spatial distributions, and spin parameters. In Section 4, we discuss what makes dark galaxies dark. Finally, we summarize our key findings in Section 5, outlining potential theoretical and practical applications for future work. To be clear, proper and comoving distances are denoted as ``Mpc/kpc" and ``cMpc/ckpc", respectively.

\section{Methodology} \label{sec:method}
To study the nature of dark galaxies, we utilize data from the IllustrisTNG simulation. This section outlines our methodology by introducing the IllustrisTNG simulation and clarifying the selection criteria for our galaxy sample.

\subsection{IllustrisTNG} \label{sec:TNG}
The IllustrisTNG (hereafter TNG) project \citep{Nelson2018, Naiman2018, Marinacci2018, Springel2018, Pillepich2018a} conducts a suite of large-volume, cosmological simulations performed with the moving-mesh code, AREPO \citep{Springel2010, Weinberger2020}. As the successor to the original Illustris project \citep{Vogelsberger2014a, Vogelsberger2014b, Genel2014, Sijacki2015}, the TNG simulations, which stands for `The Next Generation', aim to model the formation and evolution of galaxies within the framework of the $\Lambda$CDM cosmology. The new sub-resolution physics and numerical improvements of TNG are described in \citet{Weinberger2017} and \citet{Pillepich2018b}.

The TNG simulations are based on the $\Lambda$CDM cosmological parameters from \citet{Planck2016}: $\Omega_{\Lambda,0} = 0.6911$, $\Omega_{m,0} = 0.3089$, $\Omega_{b,0} = 0.0486$, $\sigma_{8} = 0.8159$, $n_{s} = 0.9667$, and $H_0 = 100\,h\,\text{km}\,\text{s}^{-1}\,\text{Mpc}^{-1}$ with $h = 0.6774$. The simulations incorporate self-gravity and hydrodynamics, as well as cosmic magnetic fields. Given the complex nature of baryonic physics and the wide range of scales involved, TNG employs various sub-resolution models to account for processes such as gas cooling, star formation, metal enrichment, and feedback mechanisms from supernova (SN) and active galactic nuclei (AGN). They also include the effects of cosmic reionization by activating a spatially uniform UV background radiation at $z \sim 6$. This cosmic reionization will be discussed in detail in Section \ref{subsec:cosmic reionization}, which turns out to be important for the formation of dark galaxies. The TNG project provides three realizations of simulations: TNG50, TNG100, and TNG300, with each number representing the side length of the simulation box in units of cMpc.

In this work, we use TNG50 \citep{Pillepich2019, Nelson2019}, as it provides the highest resolution for analyzing individual galaxies. We obtain the snapshots containing full particle data and group catalogs containing information on galaxy groups/clusters and individual galaxies at several selected redshifts covering $0 \leq z \leq 11$. We also use the galaxy merger trees constructed with SUBLINK \citep{Gomez2015} and LHaloTree \citep{Springel2005}. Because most galaxies are first identified after $z \sim 11$ on their merger trees, we trace them until $z=11$.

\begin{figure}
    \centering
    \includegraphics[width=\linewidth]{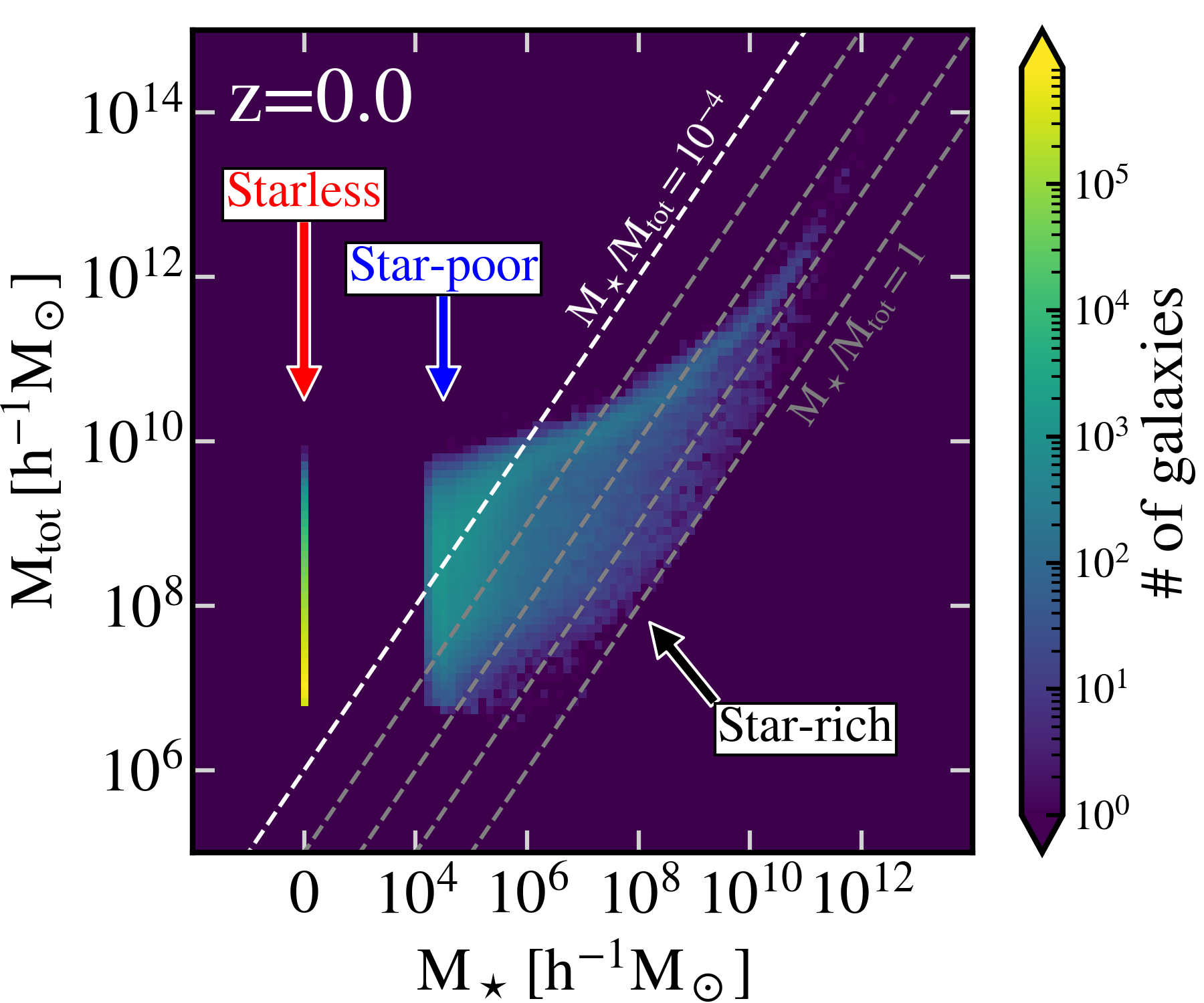}
    \caption{2D histogram of galaxies at $z=0$ in terms of their stellar masses and total masses, showing the classification of galaxies. The color indicates the number of galaxies within each bin. A white dashed line is a criterion for distinguishing luminous and dark galaxies. The gap between star-poor and starless galaxies is a result of the resolution. The minimum mass of star particle is around $10^4 \, h^{-1}$$\rm M_{\odot}$ at $z=0$.}
    \label{fig:sample_selection}
\end{figure}

A group catalog at each snapshot includes two types of groups: FoF (friends-of-friends) and SUBFIND groups. FoF groups, corresponding to galaxy groups/clusters,  are initially identified using a friends-of-friends algorithm \citep{Davis1985}. This algorithm connects particles into a single FoF group if their separation is less than the linking length, denoted as $b$ times the mean inter-particle distance. Here, $b$ represents the percolation parameter with a specific value of 0.2.

For each FOF group, the SUBFIND algorithm \citep{Springel2001, Dolag2009} is employed to identify gravitationally self-bound substructures. Such structures are identified in overdense regions in the FoF group, and then gravitationally unbound particles are removed. We use these SUBFIND groups (hereafter galaxies) that initially form as centrals in their own FoF groups and evolve on unbroken branches of merger trees.

We note that all measurements in this work are obtained from the SUBFIND group catalog. Therefore, when we use the term `halo', we are specifically referring to the self-bound structures at the scale of individual galaxies, rather than larger group or cluster-scale halos.

\begin{figure}
    \centering
    \includegraphics[width=\linewidth]{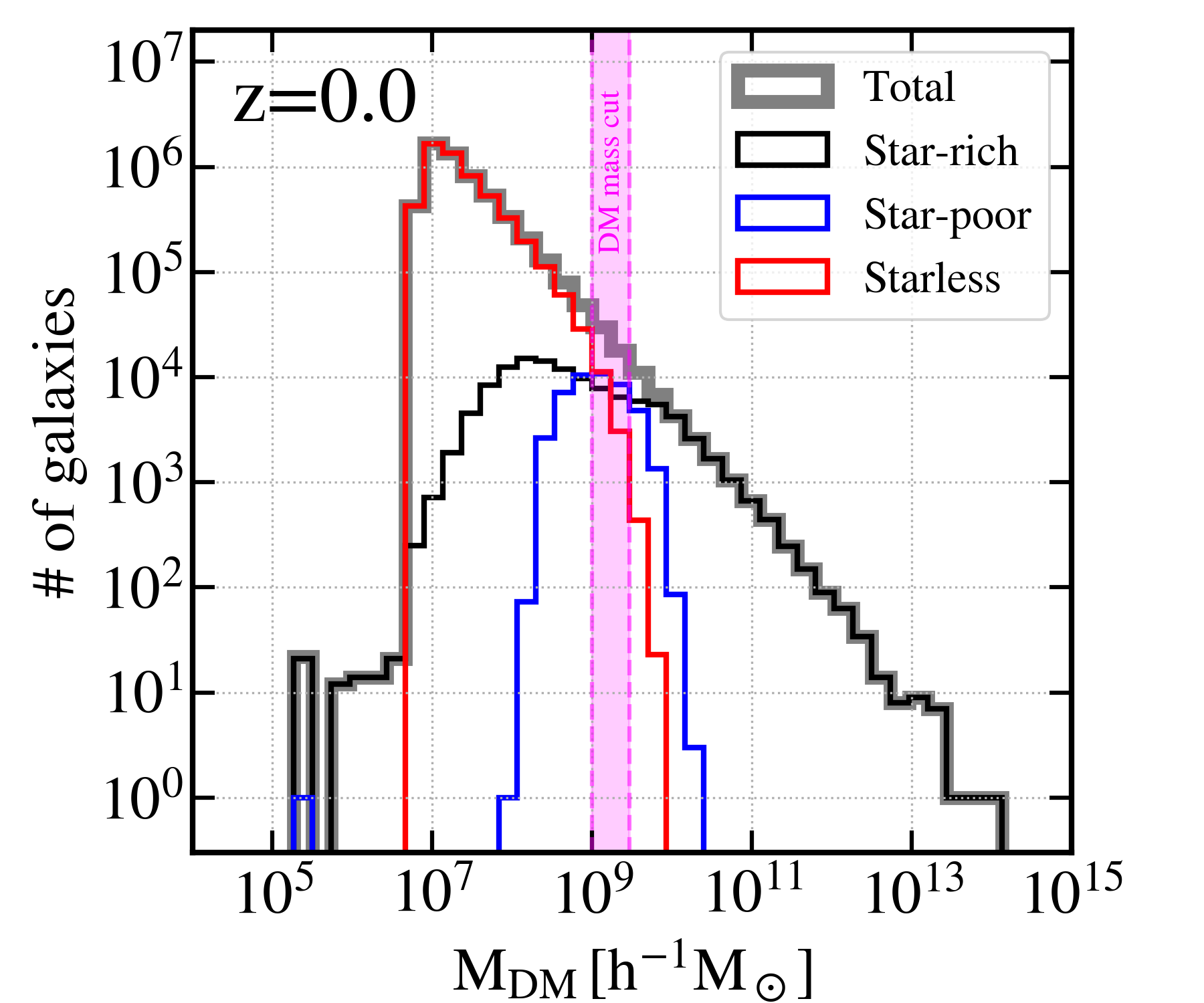}
    \caption{Dark matter mass histogram at $z=0$. The grey, black, blue, and red colors correspond to the total, star-rich, star-poor, and starless galaxies, respectively. The magenta region highlights the selection area for mass completeness. Our final galaxy sample contains 14,206 star-rich galaxies, 19,245 star-poor galaxies, and 14,318 starless galaxies.}
    \label{fig:mass_cut}
\end{figure}

\subsection{Galaxy Sample Selection} \label{sec:sample}
We classify galaxies into two groups, luminous or dark galaxies, using their stellar-to-total mass ratios at redshift $z=0$ (see Figure \ref{fig:sample_selection}). Luminous galaxies are defined as those with stellar masses greater than or equal to 1/10,000 of their total masses (i.e., $M_*/M_{\text{tot}} \geq 0.0001$), while dark galaxies have stellar masses below this threshold (i.e., $M_*/M_{\text{tot}} < 0.0001$). Our choice of this mass ratio threshold is motivated by observational studies \citep[e.g., Fig. 4 of][]{vanDokkum2018}, as we aim to focus on the objects that are darker than those typically observed. Minor adjustments to this threshold are unlikely to significantly impact our results.

\begin{figure*}
    \centering
    \includegraphics[width=1.0\linewidth]{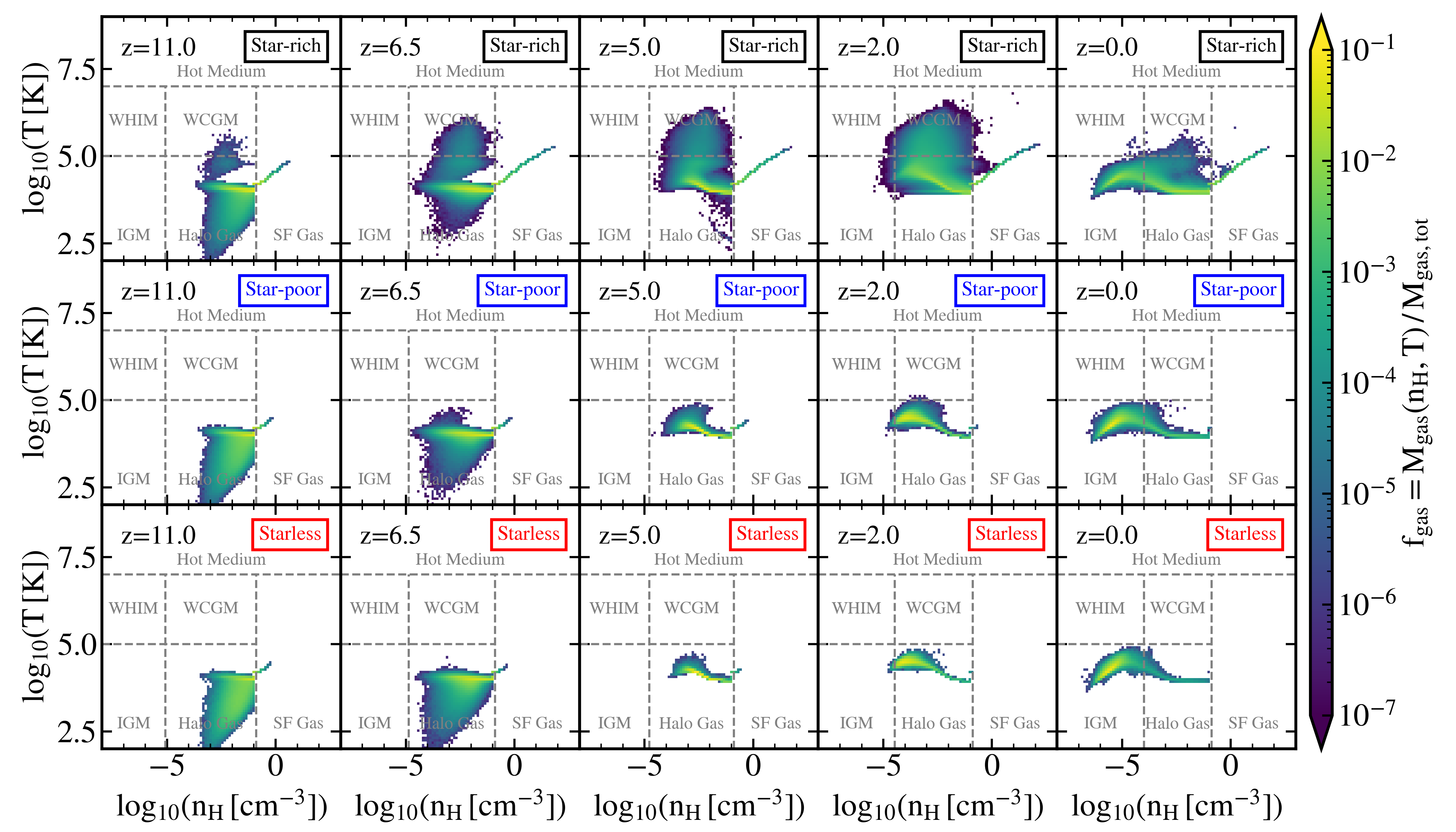}
    \caption{Gas phase diagrams in a density-temperature plane, showing the gas distributions within different types of galaxies (star-rich, star-poor, and starless) across various epochs ($z=11, 6.5, 5, 2$, and $0$). The color represents the mass fraction of gas cells in each bin. The cosmic reionization occurs at $z=6$.}
    \label{fig:gas_phase}
\end{figure*}

Dark galaxies are further divided into two groups: star-poor galaxies, which have a positive stellar mass, and starless galaxies, which have a stellar mass of zero. It should be clarified that we still call starless objects as galaxies, even though they do not have any visible stars. Similarly, we use the terms of `luminous galaxies' or `star-rich galaxies' when referring to the galaxies not classified as dark galaxies depending on the context: i.e. `luminous' vs. `dark', `star-rich' vs. `star-poor' and `starless'.

Figure \ref{fig:sample_selection} shows our classification of galaxies based on their stellar masses (x-axis) and total masses (y-axis). A white dashed line corresponds to the threshold where $M_\star/M_{\text{tot}} = 0.0001$, which separates dark galaxies from luminous galaxies. In the TNG50 group catalog, there is a total of 5,686,764 galaxies. Of these, only 116,200 galaxies are classified as luminous galaxies, while the majority of galaxies are classified as dark galaxies in our definition. The dark galaxies are composed of 47,270 star-poor galaxies and 5,524,844 starless galaxies.

Figure \ref{fig:mass_cut} represents the mass distribution of the dark matter component for each sample. The total dark matter mass distribution shows an abundance of galaxies with lower-mass dark matter halo and fewer galaxies with higher-mass dark matter halo. This trend is consistent with the initial power spectrum predicted by the $\Lambda$CDM cosmology. Luminous galaxies are prevalent at the high-mass end, while dark galaxies, particularly starless galaxies, are more abundant at the lower-mass end. The large number of starless galaxies in the low-mass range would be partly because of simulation resolution coupled with the star formation recipe. For example, the star particles do not form below a mass resolution in simulations, which can contribute to the increase in the number of starless galaxies. Additionally, the specific choice of star formation recipe can further impact the modeling of star formation in low-mass galaxies \citep{Federrath2012, Nunez2021}.

To ensure a fair comparison between luminous and dark galaxies, we select only the galaxies with dark matter mass in the range of $1.0 \times 10^9 \, h^{-1} \text{M}_\odot \leq M_{\text{DM}} \leq \, 3.0 \times 10^9 \,  h^{-1} \text{M}_\odot$. The magenta region in Figure \ref{fig:mass_cut} corresponds to this dark matter mass selection criterion. We adopt this criterion for two reasons. First, at this mass range, the three types of galaxies (star-rich, star-poor, and starless galaxies) have the most overlap in frequency, indicating similar sample sizes. Second, the selection allows us to investigate variations in baryonic content among galaxies by fixing the dark matter mass of the galaxies at $z=0$.

To summarize, our final galaxy sample includes 14,206 star-rich galaxies, 19,245 star-poor galaxies, and 14,318 starless galaxies. When we explore other redshifts, we continue tracking these galaxies across time. Hence, in this work, the type of galaxy is always referenced to their state at $z=0$.

\section{Results} \label{sec:results}
In the following section, we present our findings on dark galaxies and compare them to luminous galaxies. Our analysis encompasses diverse galaxy characteristics, such as gas properties, internal properties, environmental factors, and spin parameters.

\subsection{Gas Properties} \label{subsec:gas}
We first investigate the gas properties of both luminous and dark galaxies. Because gas is the fuel of star formation, this investigation may enable us to find hints as to why dark galaxies fail to form stars.

\subsubsection{Gas Phase Diagram}
We group the gas cells into six phases based on their temperature ($T$) and hydrogen number density ($n_{\text{H}}$), following the classification in \citet{Martizzi2019}. These phases include: hot medium, warm-hot intergalactic medium (WHIM), warm circumgalactic medium (WCGM), diffuse intergalactic medium (IGM), halo gas, and star-forming (SF) gas. Our primary interest is in the SF gas phase, identified in the TNG simulations using a specified density threshold of $n_{\text{H}} = 0.13 \, \text{cm}^{-3}$. The gas cells with density higher than this threshold are stochastically converted into star particles on a timescale determined by the local cold gas density \citep{Springel2003}.

Figure \ref{fig:gas_phase} displays gas phase diagrams for star-rich, star-poor, and starless galaxies at different epochs. Each row corresponds to a distinct galaxy type, while each column represents a specific redshift. The color of each bin indicates the mass fraction of the gas within that bin. The mass fraction is the gas mass in each bin divided by the total gas mass for a given galaxy type.

The most prominent feature in Figure \ref{fig:gas_phase} is the different amount of SF gas between luminous and dark galaxies. Star-rich galaxies consistently exhibit substantial amounts of SF gas throughout cosmic time. In contrast, star-poor and starless galaxies show minimal or negligible amounts of SF gas. The rightmost column, representing the present epoch ($z=0$), clearly shows the abundance of SF gas in star-rich galaxies, whereas star-poor and starless galaxies show a noticeable absence of SF gas. This trend persists even in the early universe, as shown in the leftmost column (at $z = 11$). These findings indicate that luminous galaxies are characterized by an initial abundance of gas reservoirs available for star formation, while dark galaxies lack such a crucial resource.

Another notable feature in Figure \ref{fig:gas_phase} is the increase of gas temperature between $z=6.5$ and $z=5$. At $z=6.5$, there exists a portion of gas with temperatures below $\rm 10^4\,K$. However, at $z=5$, the majority of gas, especially in dark galaxies, have temperatures exceeding $\rm10^4\,K$. This temperature transition is closely associated with the cosmic reionization realized in the TNG simulation. The cosmic reionization in TNG50 starts to heat up gas after $z=6$ based on the assumption of a spatially uniform UV background radiation. We will provide more detailed information on the numerical implementation of cosmic reionization in the TNG simulation in Section \ref{subsec:cosmic reionization}.

Additionally, Figure \ref{fig:gas_phase} represents that luminous galaxies contain warm gas of $T > 10^5$K (e.g., WHIM and WCGM) throughout cosmic history. On the other hand, dark galaxies have a lack of such warm gas. The presence of warm gas in luminous galaxies can be attributed to the by-products of SN feedback. As the global star formation rate is regulated after cosmic noon ($z < 2$), the warm gas in luminous galaxies gradually cools down by $z=0$. Figure \ref{fig:gas_phase} also shows that none of the galaxies in our sample contain hot medium ($T > 10^7$K) due to low mass of the galaxy sample.

\begin{figure}
    \centering
    \includegraphics[width=\linewidth]{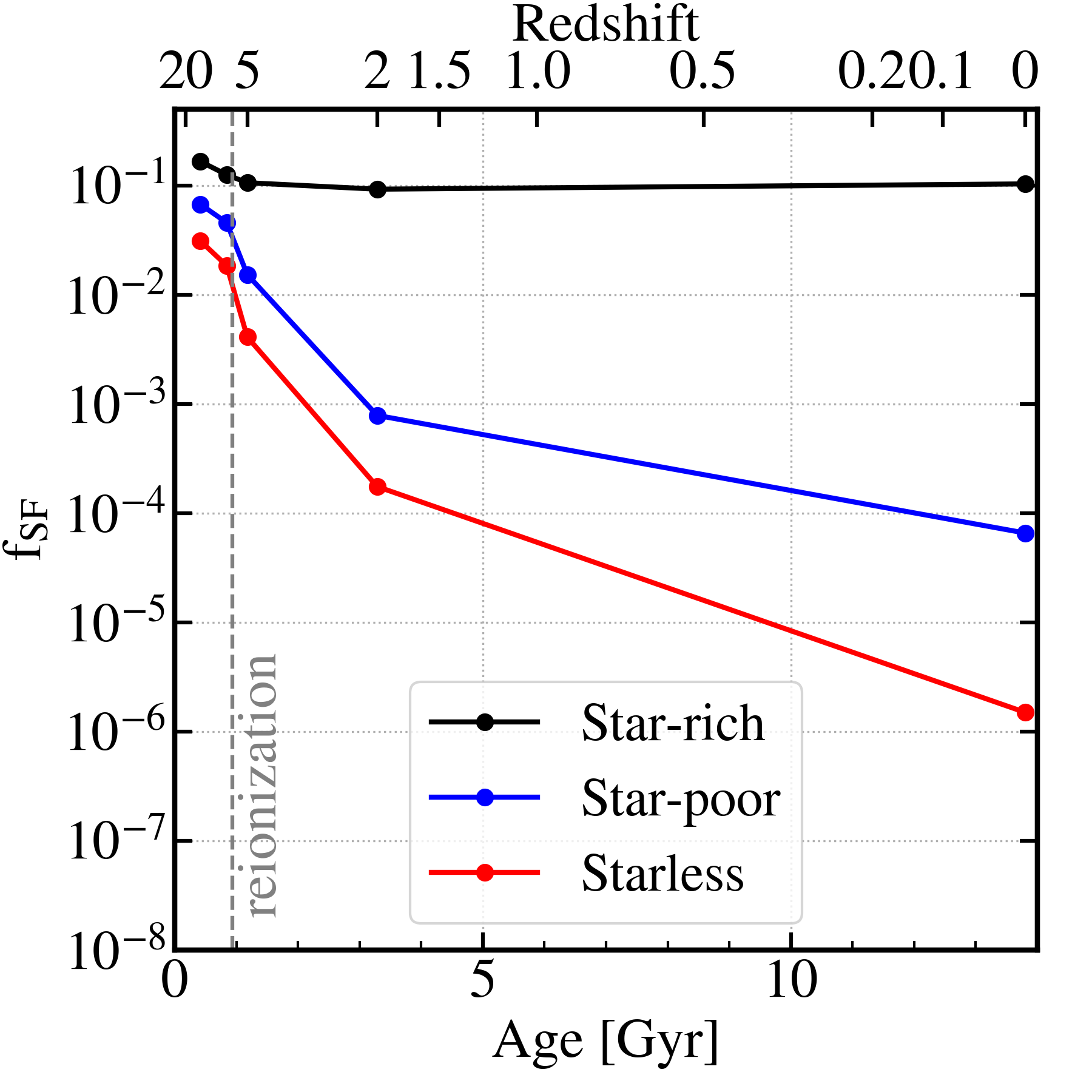}
    \caption{Star-forming gas fractions as a function of time for star-rich, star-poor, and starless galaxies. The plot shows that star-poor and starless galaxies consistently have lower star-forming gas fractions than star-rich galaxies, with a notable decrease occurring after cosmic reionization ($z=6$). In contrast, star-rich galaxies maintain a relatively constant star-forming gas fraction throughout cosmic history.}
    \label{fig:gas_farction}
\end{figure}

\begin{figure*}
    \centering
    \includegraphics[width=1.0\linewidth]{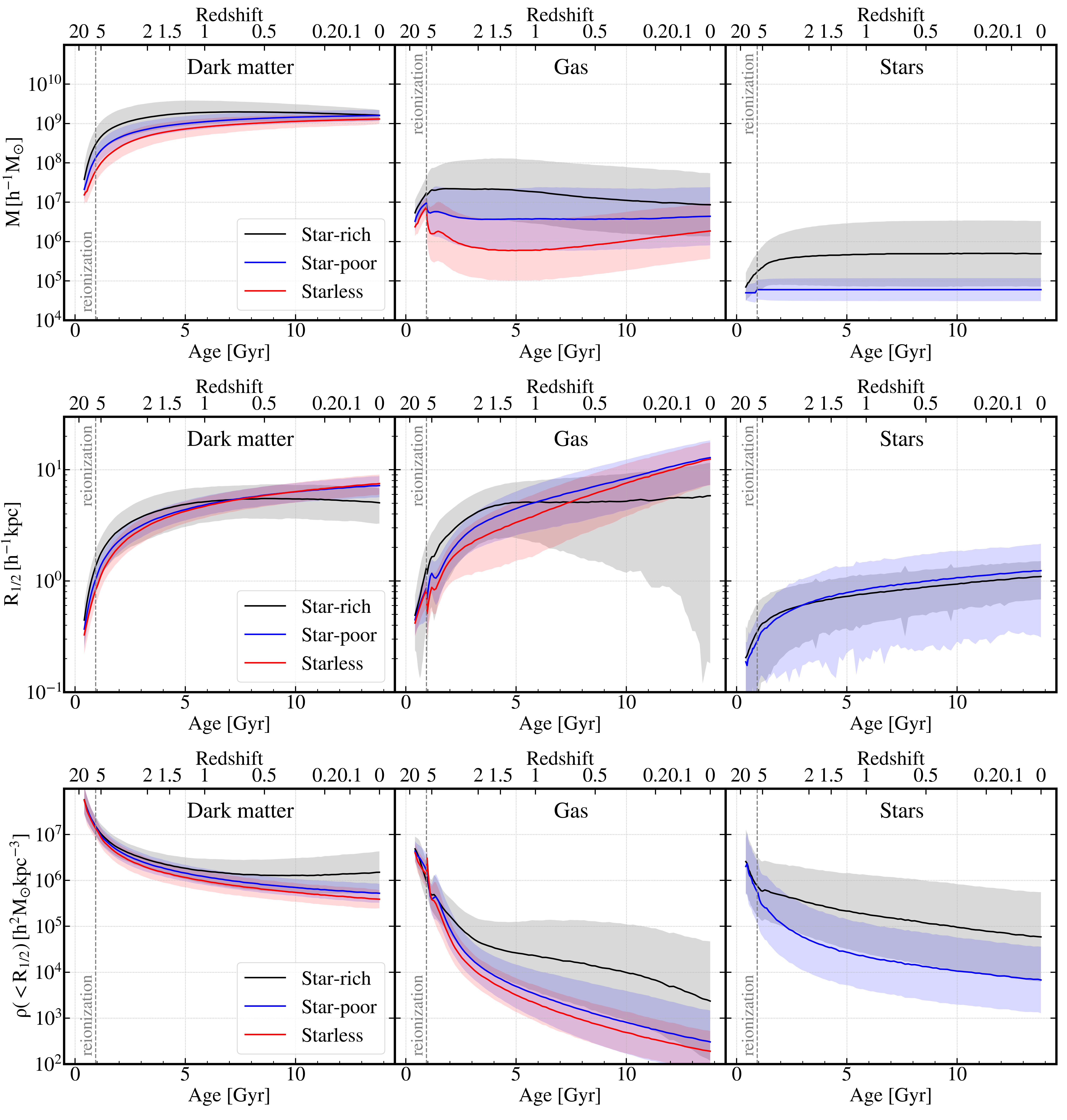}
    \caption{Evolutionary history of star-rich, star-poor, and starless galaxies, in terms of mass (top row), size (half-mass radius; middle row), and density (bottom row). Each row consists of three panels, addressing dark matter on the left, gas in the middle, and stellar components on the right. The curves represent the median values with the shaded regions indicating the 1-$\sigma$ scatter. Note that the curves and 1-$\sigma$ scatters are measured from the galaxies with positive stellar or gas mass.}
    \label{fig:evolution}
\end{figure*}

\subsubsection{Star-forming Gas Fraction}
We now shift our focus to the star-forming gas, given its crucial role in galaxy evolution. To quantify its prevalence, we calculate the mass fraction of star-forming gas over time using the equation:
\begin{equation}
    f_{\text{SF}} = M_{\text{gas, SF}} / M_{\text{gas,tot}}\text{,}
\end{equation}
where $M_{\text{gas, SF}}$ and $M_{\text{gas,tot}}$ denote the masses of star-forming gas and of total gas for the specific galaxy type, respectively.

Figure \ref{fig:gas_farction} shows the SF gas fraction as a function of time. It is evident that dark galaxies consistently have lower SF gas fractions than luminous galaxies, even in the early universe at $z=11$. Notably, the SF gas fractions in dark galaxies undergo a significant decrease after cosmic reionization. The cosmic reionization interrupts gas cooling, which in turn hinders the collapse of gas and eventually suppresses the SF gas fraction in dark galaxies. After the epoch of cosmic reionization ($z<6$), dark galaxies suffer from the decrease of SF gas all the way down to $z=0$, while luminous galaxies maintain a relatively constant SF gas fraction throughout cosmic history. This result shows that luminous galaxies have survived from the effect of cosmic reionization and retained a substantial amount of SF gas.

\subsection{Internal Properties} \label{subsec:internal}
Using the merger tree of TNG50, we trace the evolutionary history of the key properties of galaxies, i.e., their mass, size, and density. Figure \ref{fig:evolution} shows the evolutionary histories of these properties. Here, we present the plots for dark matter and baryonic components (gas and star) separately. Each curve on the plot represents the median value with the shaded region indicating the 1-$\sigma$ scatter. We note that the curves and 1-$\sigma$ scatters are measured from the galaxies with positive stellar or gas mass.

\subsubsection{Mass} \label{subsubsec:Mass}
The top row of Figure \ref{fig:evolution} shows the evolutionary history of mass. The left panel corresponds to dark matter mass history. The dark matter mass increases steadily over time from around $10^7 \, h^{-1} M\odot$ to $10^9 \, h^{-1} M\odot$. Because we applied the dark matter mass criteria when selecting the galaxy samples (as described in Section \ref{sec:sample}), the final dark matter masses of all types of galaxies are expected in a similar range. Nonetheless, we find that the dark matter mass of luminous galaxies increases faster than that of dark galaxies. Specifically, it takes 1.2 Gyr for star-rich galaxies to reach half of their final dark matter mass, while star-poor and starless galaxies need 2.2 Gyr and 3.5 Gyr, respectively. This suggests that luminous galaxies have experienced more efficient accretion and merging events in earlier cosmic epochs. We could notice another interesting feature that is a slight decrease in dark matter within luminous galaxies. This reduction can be attributed to the process of stripping, where dark matter is gradually pulled into larger neighboring halos.

The middle panel shows the evolutionary history of gas mass. Initially, luminous galaxies start with slightly larger gas mass than dark galaxies. Around 1 Gyr, both star-poor and starless galaxies experience a substantial reduction in their gas mass. This decrease is attributed to cosmic reionization, which leads to gas heating and ejection from halos. Notably, this effect is much less pronounced in star-rich galaxies, allowing them to retain a higher fraction of gas reservoirs.

The right panel shows the evolutionary history of stellar mass. Luminous galaxies undergo star formation by consuming the gas before the cosmic noon ($z > 2$), and they retain their stars over time. In contrast, star-poor galaxies have only a few star particles and do not produce additional stars. As for starless galaxies, a few of them initially possess a small number of star particles; they then lose the star particles and fail to form new ones. The stellar mass history of starless galaxies is not shown here.

\begin{figure*}
\centering
    \begin{tabular}{cc}
         \includegraphics[width=0.48\textwidth]{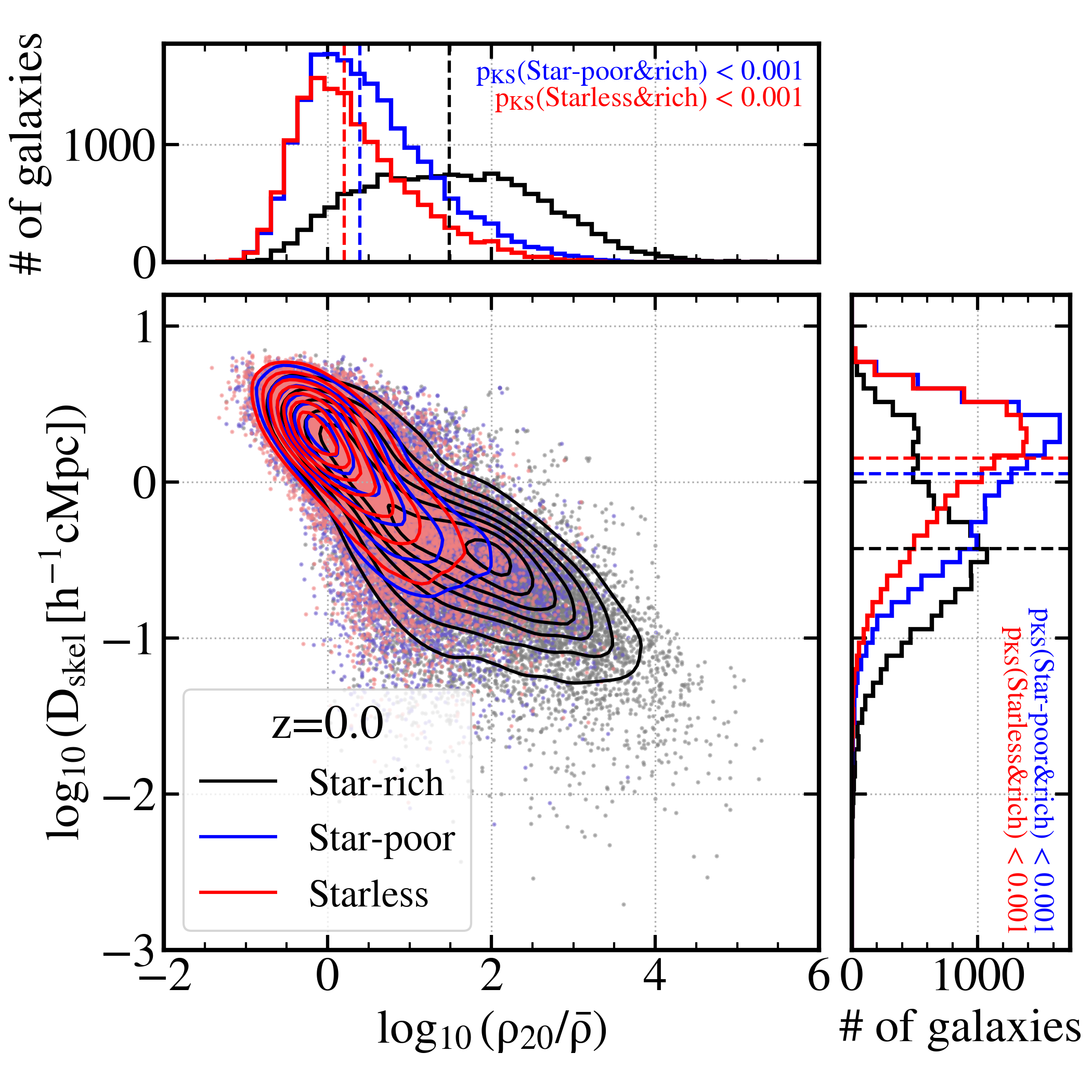}&
         \includegraphics[width=0.48\textwidth]{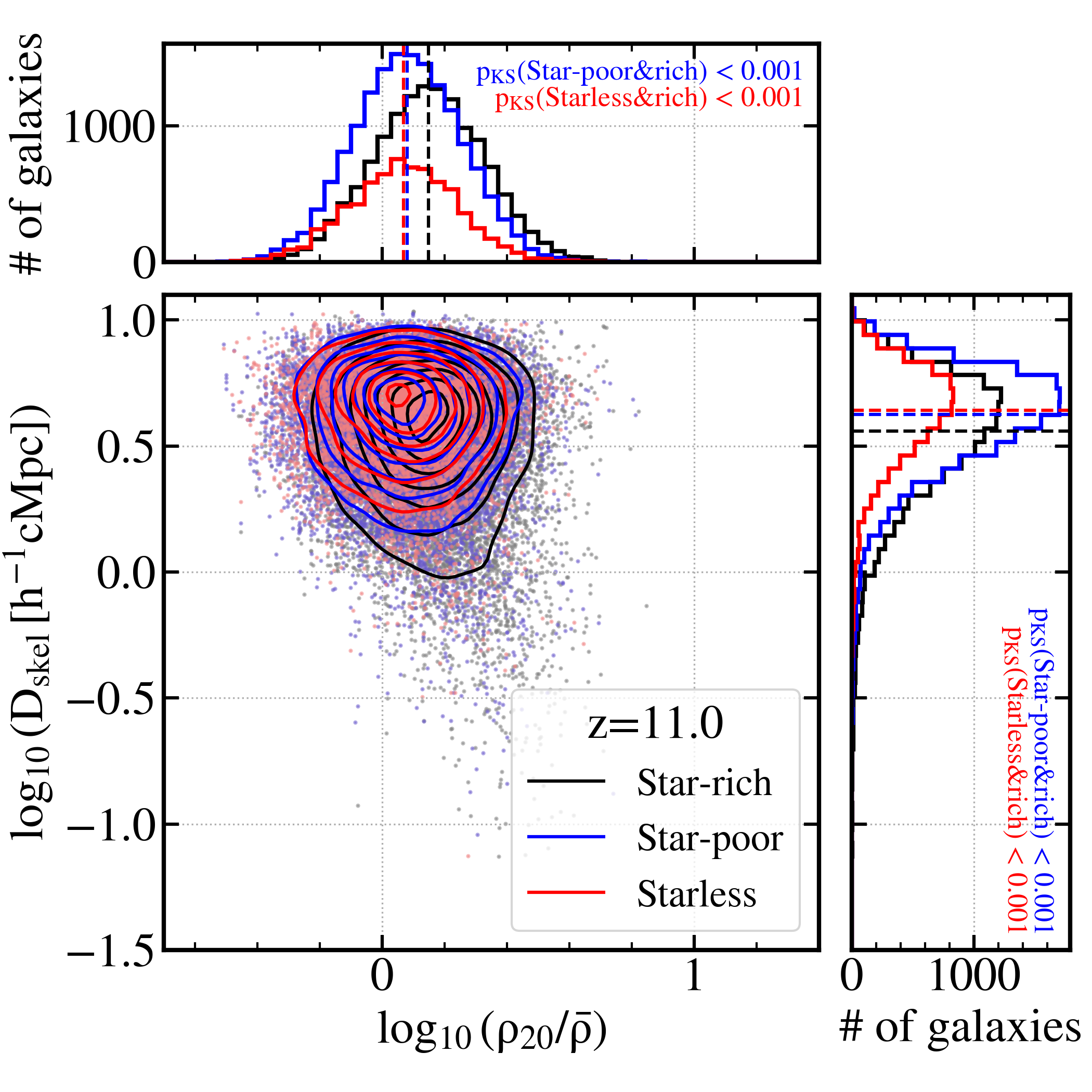}
    \end{tabular}
    \caption{Spatial distribution of galaxies based on local density and distance to the nearest filament. The black, blue, and red dots represent star-rich, star-poor, and starless galaxies, respectively, while the contours show their distribution of them. \textit{Left}: Spatial distribution of galaxies at $z=0$. Star-rich galaxies tend to be closer to filaments and located in denser regions, while dark galaxies are more distant from filaments and found in less dense regions. \textit{Right}: Spatial distribution at $z=11$. Note that the x and y ranges are different, with a narrower range than the left panel to visualize the details better. Although the contours are closer together due to less developed large-scale structures, an offset remains between luminous and dark galaxies. This suggests distinct origins with potential implications for their future evolution. The p-values from the Kolmogorov-Smirnov test indicate statistically significant differences between the distributions of luminous and dark galaxies.}
\label{fig:spatial}
\end{figure*}

\subsubsection{Size}
To investigate the spatial extent of dark matter and baryonic components, we analyze their half-mass radius, also referred to as size in this study. The middle row of Figure \ref{fig:evolution} shows the evolutionary history of the size of dark matter, gas, and stellar components within the galaxies.

The left panel corresponds to the evolutionary trend in the size of dark matter halos within galaxies. In the early universe, the growth rate of dark matter halos sizes within luminous galaxies surpasses that of dark galaxies, in accordance with their growing mass. However, after 7.5 Gyr, dark matter halos in luminous galaxies undergo a slight contraction, while those in dark galaxies continue to expand. As a result, at $z=0$, dark galaxies exhibit more extended dark matter distribution than luminous galaxies.

The middle panel represents the evolutionary history of gas size. At $z=6$, there are fluctuations of gas size influenced by cosmic reionization. After cosmic noon ($z < 2$), luminous galaxies tend to maintain their gas size, contrasting with dark galaxies that continue expanding in gas size. This difference could be because luminous galaxies can transfer their angular momentum through interactions with other galaxies, while dark galaxies are in isolation. We will discuss more about it in Section \ref{subsec:mergers_interactions}.

Lastly, the right panel shows the evolutionary history of stellar size. Both star-rich and star-poor galaxies exhibit a gradual increase in their stellar size over time, with no significant differences between them.

\subsubsection{Density}
The density is computed by dividing their half-mass by the spherical volume encompassing the half-mass radius. The bottom row of Figure \ref{fig:evolution} shows the evolution of the density of dark matter, gas, and stellar components for the different types of galaxies.

In the left panel, the dark matter density of all galaxies decreases over time, which is consistent with their expanding sizes. However, the dark matter density in luminous galaxies decreases at a slower rate than in dark galaxies. Consequently, at $z=0$, luminous galaxies exhibit denser dark matter halos than dark galaxies.

The middle panel shows that the gas densities of luminous and dark galaxies are similar in the early universe ($z>6$). However, over time, the gas densities within dark galaxies become lower than those in luminous galaxies. This decrease in gas densities in dark galaxies correlate with their decreasing masses and expanding sizes. The reduction in gas density could suppress star formation in dark galaxies.

The right panel shows the evolution of stellar density. Initially, both star-rich and star-poor galaxies have similar stellar densities. However, as time goes on, star-poor galaxies exhibit a lower stellar density than star-rich galaxies.

\subsection{Spatial Distribution} \label{subsec:environment}
To explore the spatial distributions of luminous and dark galaxies, we analyze their local densities ($\rho_{20}$) as well as their distances to the nearest filament ($D_{\text{skel}}$).

The local density indicates how crowded galaxy neighborhoods are. For a given galaxy, we calculate the local density by using surrounding dark matter particles as tracers. Because there are a huge number of dark matter particles in the full snapshot data ($\# = 10,077,696,000$), we choose a smaller subset by random resampling. This subset contains 0.001\% of the total dark matter particles. With this resampled dataset, we compute the local dark matter density using the following equation:
\begin{equation}
    \rho_{n} = \sum^{n}_{i=1} W(|\bold{r}_i|, h_{\text{sml}}).
\end{equation}
Here, $n$ is the number of neighboring dark matter particles, while $W(|\bold{r}_i|, h_{sml})$ is the kernel function to smooth out the distribution of dark matter particles. Additionally, $\bold{r}_i$ denotes the distance between a given galaxy and the $i$-th dark matter particle, and $h_{\text{sml}}$ represents the smoothing length, corresponding to the distance to a $n$-th dark matter particle. We choose $n=20$, because the distance to the 20th nearest dark matter particle corresponds to a few Mpc, which is suited for the consideration of large-scale structural properties \citep{Park2007, Muldrew2012}. Furthermore, we normalize the $\rho_{20}$ value by estimating the mean mass density of dark matter over the entire simulation box, denoted as $\bar{\rho}$.

We also compute the distance of each galaxy to its nearest filament. To extract the filamentary structures from the resampled distribution of dark matter particles, we use the Discrete Persistent Structure Extractor code \citep[DisPerSE, ][]{Sousbie2011a, Sousbie2011b}. We adopt a minimum persistence level of 6$\sigma$, ensuring that the selected filamentary structures are distinct from random fluctuations at a confidence level of 6$\sigma$. For each galaxy, we identify the closest filament and measure the distance to it, denoted as $D_{\text{skel}}$.

Figure \ref{fig:spatial} represents the spatial distributions of galaxies in terms of these two parameters: 
$\rho_{20}/\bar{\rho}$ (x-axis) and $D_{\text{skel}}$ (y-axis), which indicate isotropic and anisotropic environments, respectively \citep[e.g.,][]{Park2007, Kraljic2018}.

The left panel of Figure \ref{fig:spatial} shows the spatial distribution of star-rich, star-poor, and starless galaxies at the present epoch ($z = 0$). Individual galaxies are denoted by dots, while contours illustrate their distributions. Star-rich, star-poor, and starless galaxies correspond to black, blue, and red contours, respectively. We realize that the center of black contours is located towards the lower right side, indicating that star-rich galaxies tend to be closer to filaments and located in denser regions. In contrast, the centers of red and blue contours are located towards the upper left side, indicating star-poor and starless galaxies tend to be far from filaments and located in less dense regions.

\begin{figure}
    \centering
    \begin{tabular}{c}
        \includegraphics[width=\linewidth]{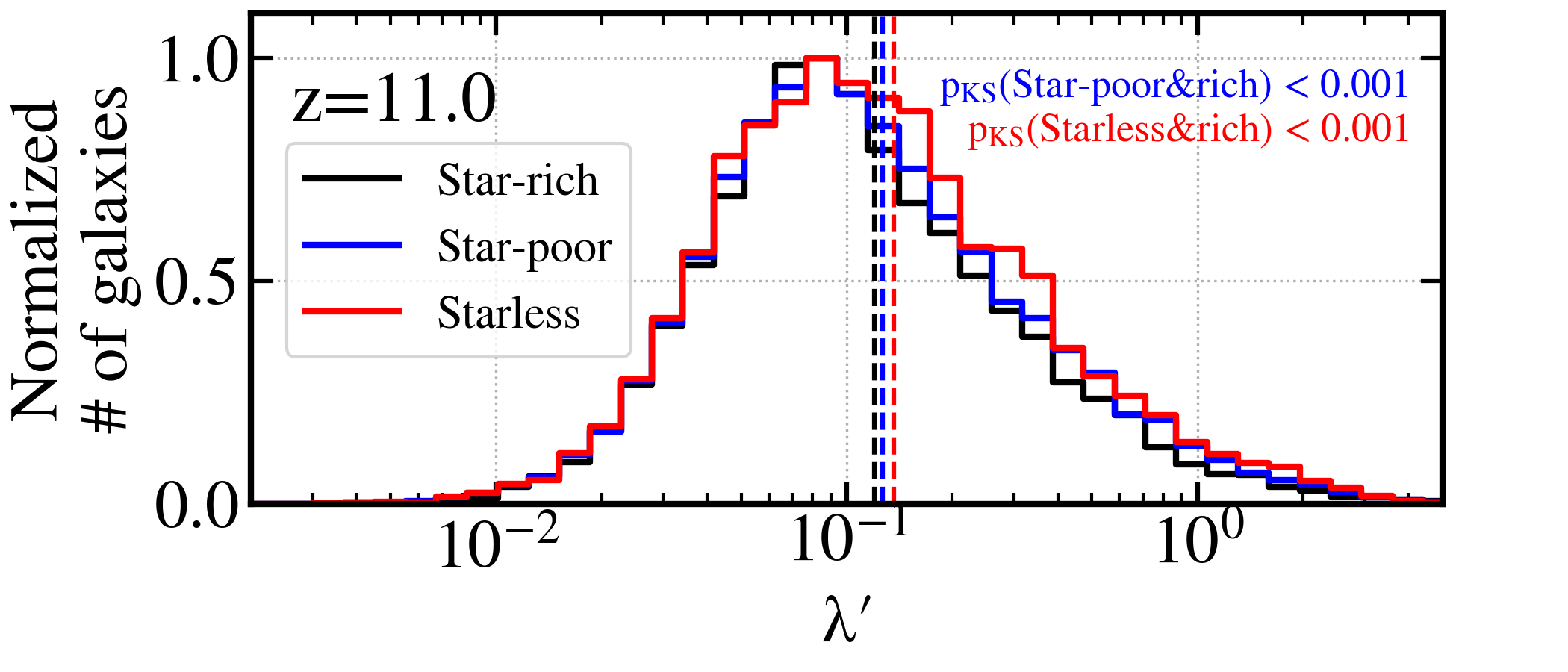} \\
        \includegraphics[width=\linewidth]{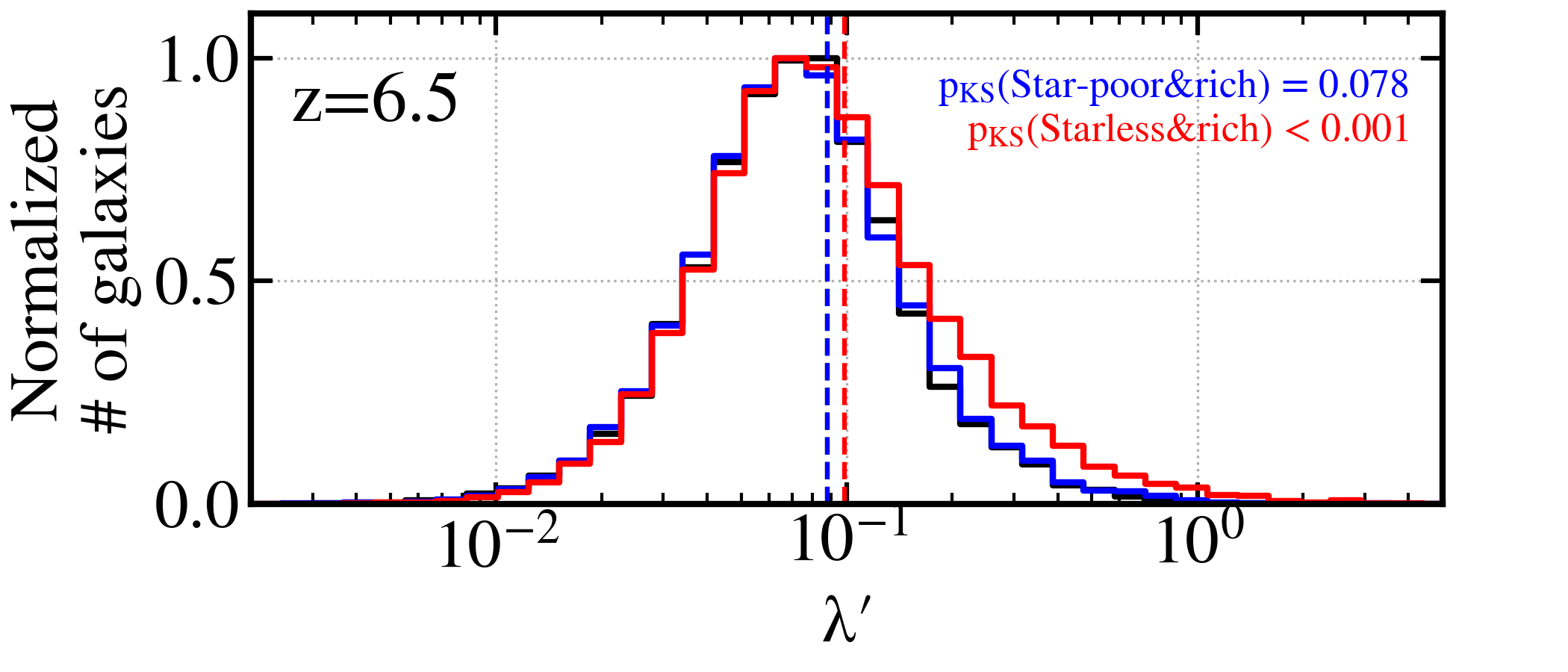} \\
        \includegraphics[width=\linewidth]{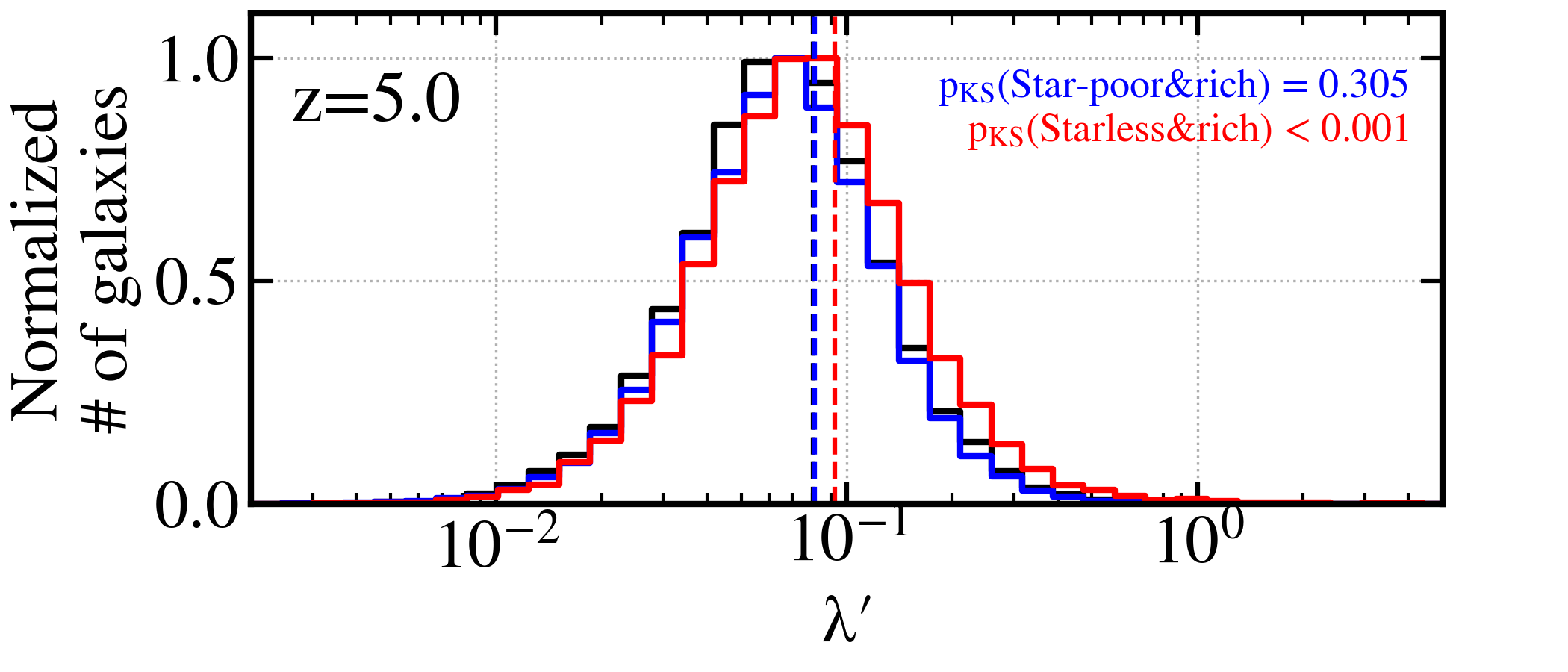} \\
        \includegraphics[width=\linewidth]{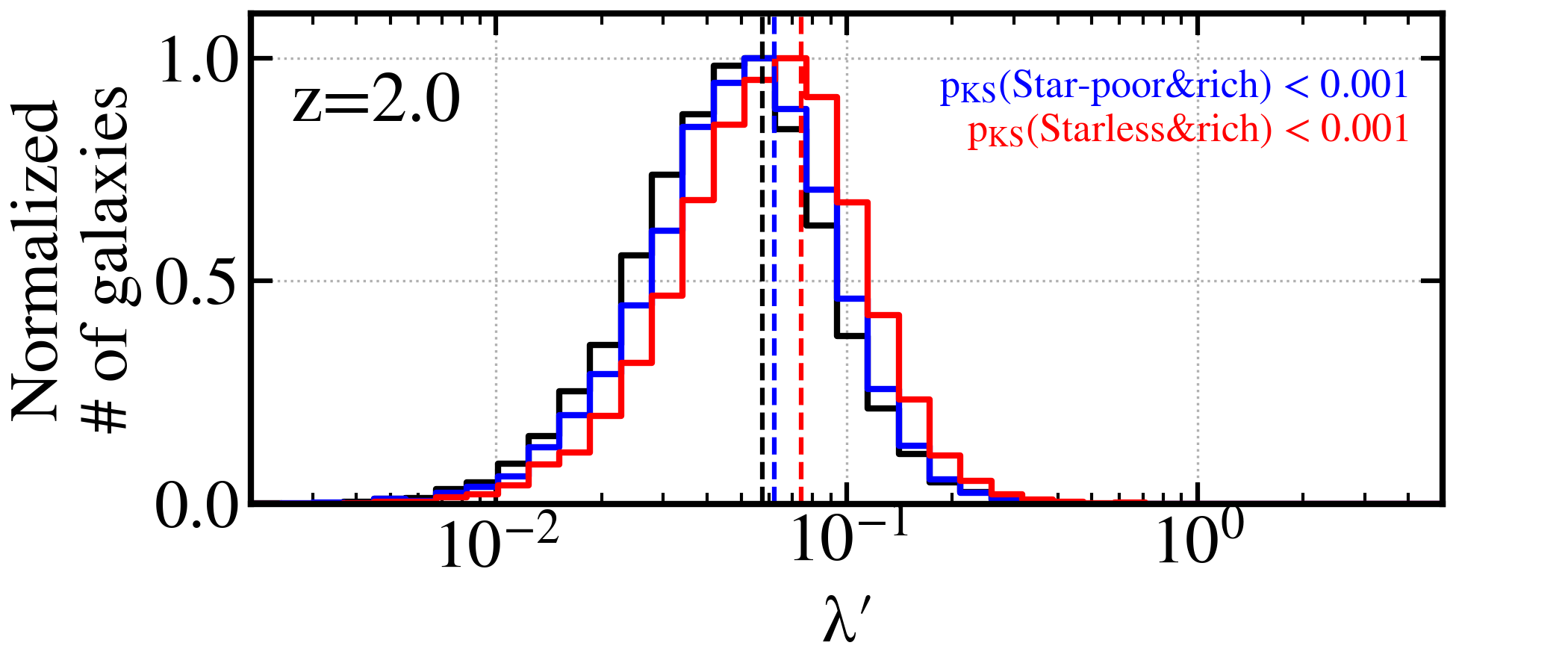} \\
        \includegraphics[width=\linewidth]{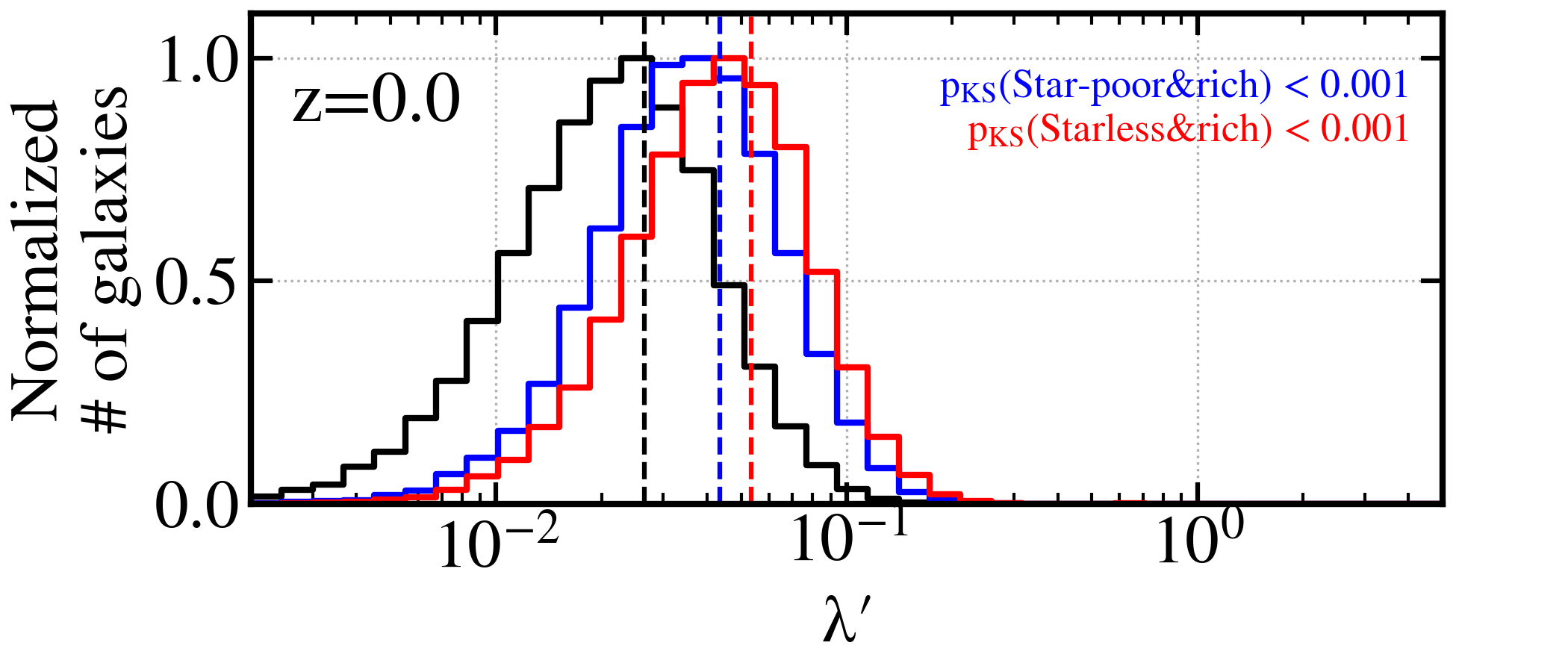}
    \end{tabular}
    \caption{Normalized histogram of the Bullock spin parameter for galaxy halos at different redshifts. From top to bottom, the redshifts change from $z=11, 6.5, 5, 2$ and $0$. The dashed line represents the median values. In the early universe (at $z=11$), there is little difference in the spin parameter between star-rich, star-poor, and starless galaxies. However, at $z=0$, star-rich galaxies exhibit a lower spin parameter than other types of galaxies.} 
    \label{fig:spin_bullock}
\end{figure}

The right panel of Figure \ref{fig:spatial} mirrors the left panel but shows the distribution in the early universe at $z = 11$, corresponding to when the universe is $\sim 0.4$ Gyr old. The contours appear closer together because large-scale structures had not yet fully developed at this point. This leads to a small difference in the distribution of the contours among the samples. Despite the close proximity of the contours, there exists an offset between lumionus and dark galaxies, which is confirmed by the Kolmogorov-Smirnov (K-S) test (see the p-values in each histogram panel). This finding implies that luminous and dark galaxies originate from slightly different locations, and these disparate origins could impact their future evolution, such as influencing gas accretion or mass growth rates. 

\subsection{Spin Parameters} \label{subsec:spin}
The spin of dark matter halo can be influenced by the tidal torque from its surrounding environment, which in turn can have implications for gas accretion. Therefore, it could be an important component in the evolution of galaxies. Here, we examine the spin parameter proposed by \citet{Bullock2001}, which is defined as
\begin{equation}
    \lambda' = \frac{|\mathbf{J}|}
                    {\sqrt{2}MRV} \text{,}
\end{equation}  
where $\mathbf{J}$ is the angular momentum of a galaxy halo, $M$ and $R$ are the virial mass and virial radius, respectively, and $V$ is the virial circular velocity, given by $V = \sqrt{GM/R}$ with the gravitational constant $G$. 

The virial radius of a halo can be described as the point where the enclosed mass density meets the criteria in the following equation:
\begin{equation}
    \rho(<R) = \Delta_{c} \rho_{c} \text{.}
\end{equation}
Here, $\rho_{c}$ is the redshift-dependent critical density of the universe, defined as
\begin{equation}
    \rho_{c}(z) = \frac{3 H^2(z)}{8 \pi G} \text{,}
\end{equation}
with the Hubble parameter $H^2(z) = h^2 [\Omega_{m,0} (1+z)^3 + \Omega_{\Lambda,0}]$. In the context of a flat universe \citep{Bryan1998}, $\Delta_{c} = 18\pi^2 + 82x - 39 x^2$ where $x = \Omega(z)-1$, and $\Omega(z)$ is the matter content at redshift $z$ defined as
\begin{equation}
    \Omega(z) = \frac{\Omega_{m,0} (1+z)^3}
                     {\Omega_{m,0} (1+z)^3 + \Omega_{\Lambda,0}} \text{.}
\end{equation}
Once measuring the virial radius, we can easily obtain the virial mass as follows:
\begin{equation}
    M = \frac{4 \pi R^3}{3} \rho(<R) \text{.}
\end{equation}

Figure \ref{fig:spin_bullock} displays histograms of the spin parameters measured with dark matter particles at different redshifts. In the bottom-most panel, it is evident that dark galaxies have larger spin parameters than luminous galaxies at $z=0$. In the top-most panel (corresponding to the highest redshift $z=11$), the distributions of spin parameters between luminous and dark galaxies overlap significantly. However, the small p-values derived from the K-S test (shown in each panel) indicate that we can reject the null hypothesis suggesting that spin parameters of luminous and dark galaxies are drawn from the same underlying distribution. Therefore, our finding indicates that dark galaxies have slightly larger spin parameters than those of luminous galaxies.

We also find a decreasing trend in the spin parameters over time for both luminous and dark galaxies. Notably, the spin parameters of luminous galaxies diminish more significantly after cosmic noon ($z < 2$). This decline in the spin of star-rich galaxies can be attributed to the transfer of angular momentum through mergers and interactions with other galaxies. A more detailed discussion about this phenomenon will be provided in Section \ref{subsec:mergers_interactions}.

\section{Discussion} \label{sec:discussion}
In this section, we delve into the origin and evolution of dark galaxies based on our key findings. Our study reveals that multiple factors can contribute to the formation of dark galaxies. Here we discuss each factor in detail.

\subsection{Early environments}
The distinct properties of present-day luminous and dark galaxies originate from their early environments, such as their birthplace and the amount of available gas for star formation (See Figures \ref{fig:gas_farction} and \ref{fig:spatial}). Our findings suggest that galaxies formed in less dense environments with a scarcity of star-forming gas are more likely to become dark galaxies. In contrast, galaxies formed in denser environments with sufficient star-forming gas are more prone to develop into luminous galaxies.

Moreover, we also find a subtle disparity in the initial spins of dark galaxies, which tend to be slightly larger than those of luminous galaxies even though the differences in spin are not significant (see Figure \ref{fig:spin_bullock}). This is consistent with the findings reported in the previous studies \citep{Jimenez1997, Kim2013, Jimenez2020}.

\subsection{Mergers and Interactions} \label{subsec:mergers_interactions}
As galaxies evolve, the differences between luminous and dark galaxies become more pronounced. We use the merger history data \citep{Gomez2017, Eisert2023}, and find that luminous galaxies in denser regions have experienced a larger number of mergers than dark galaxies in sparser regions. 
Here, it should be noted that the merger history data is constructed for the cases when galaxies have non-zero stellar mass. Therefore, galaxies with no star particles (i.e., starless galaxies in this study) are not properly considered in this analysis.

\begin{figure}
    \centering
    \includegraphics[width=\linewidth]{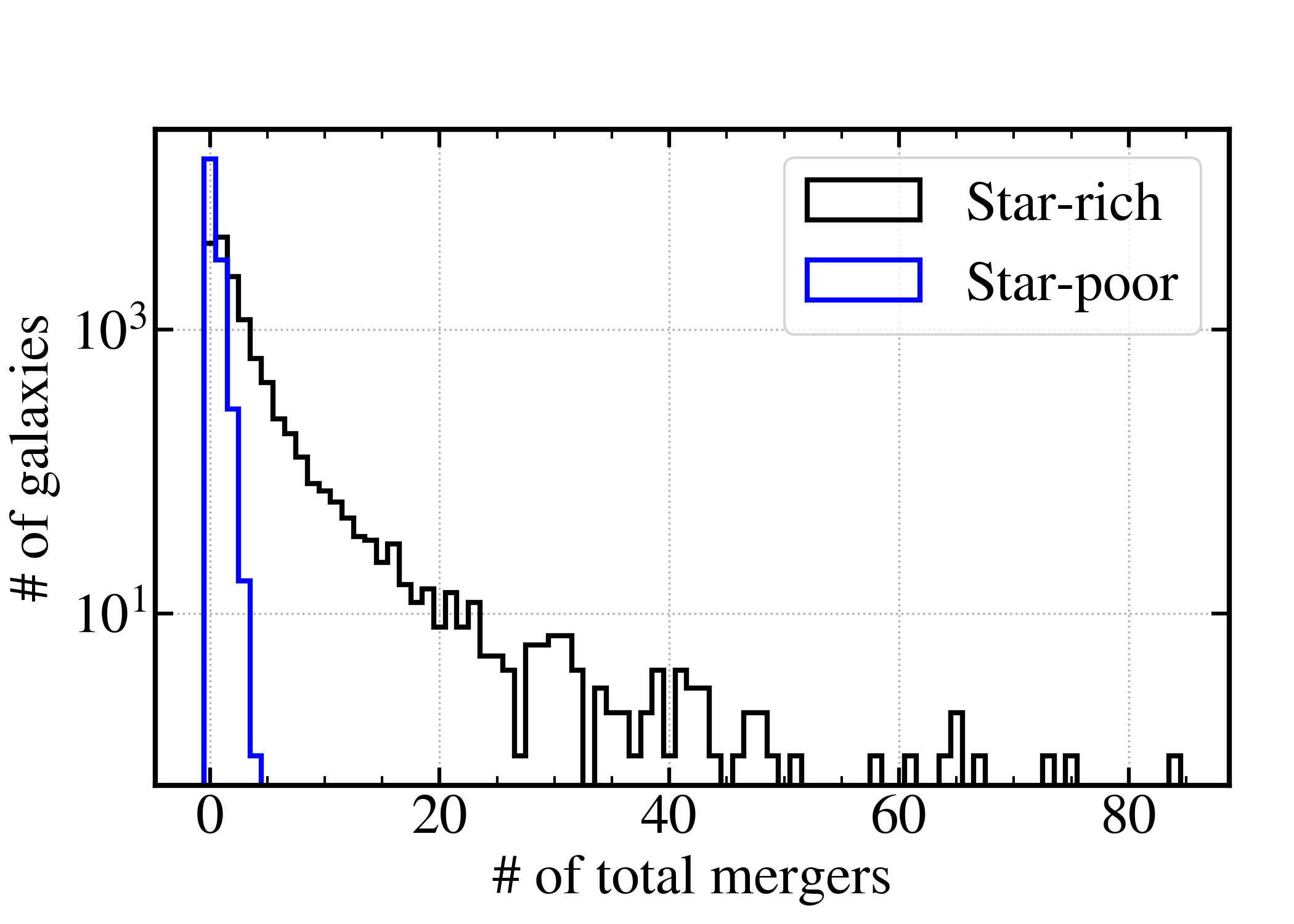}
    \caption{Merger history of star-rich and star-poor galaxies. The histogram shows the total number of mergers experienced by galaxies until $z=0$. Note that only mergers involving stellar systems are considered here.}
    \label{fig:merger}
\end{figure}

Figure \ref{fig:merger} shows the total number of mergers across the cosmic history. It highlights that star-rich galaxies have experienced numerous mergers with the systems containing stars, but star-poor galaxies have largely remained unaffected. As noted above, the plot shows the cases when the mergers occur between galaxies with stars. If there are galaxies that may have merged with starless systems, the data in this plot do not reflect such cases.

We also use the merger tree to investigate the satellite fraction ($f_{\text{sat}}$) as a function of time for luminous and dark galaxies. The satellite fraction represents the proportion of luminous or dark galaxies that are satellites of other, more massive galaxies. This is computed by dividing the number of satellite luminous or dark galaxies by the total number of galaxies.

\begin{figure}
    \centering
    \includegraphics[width=\linewidth]{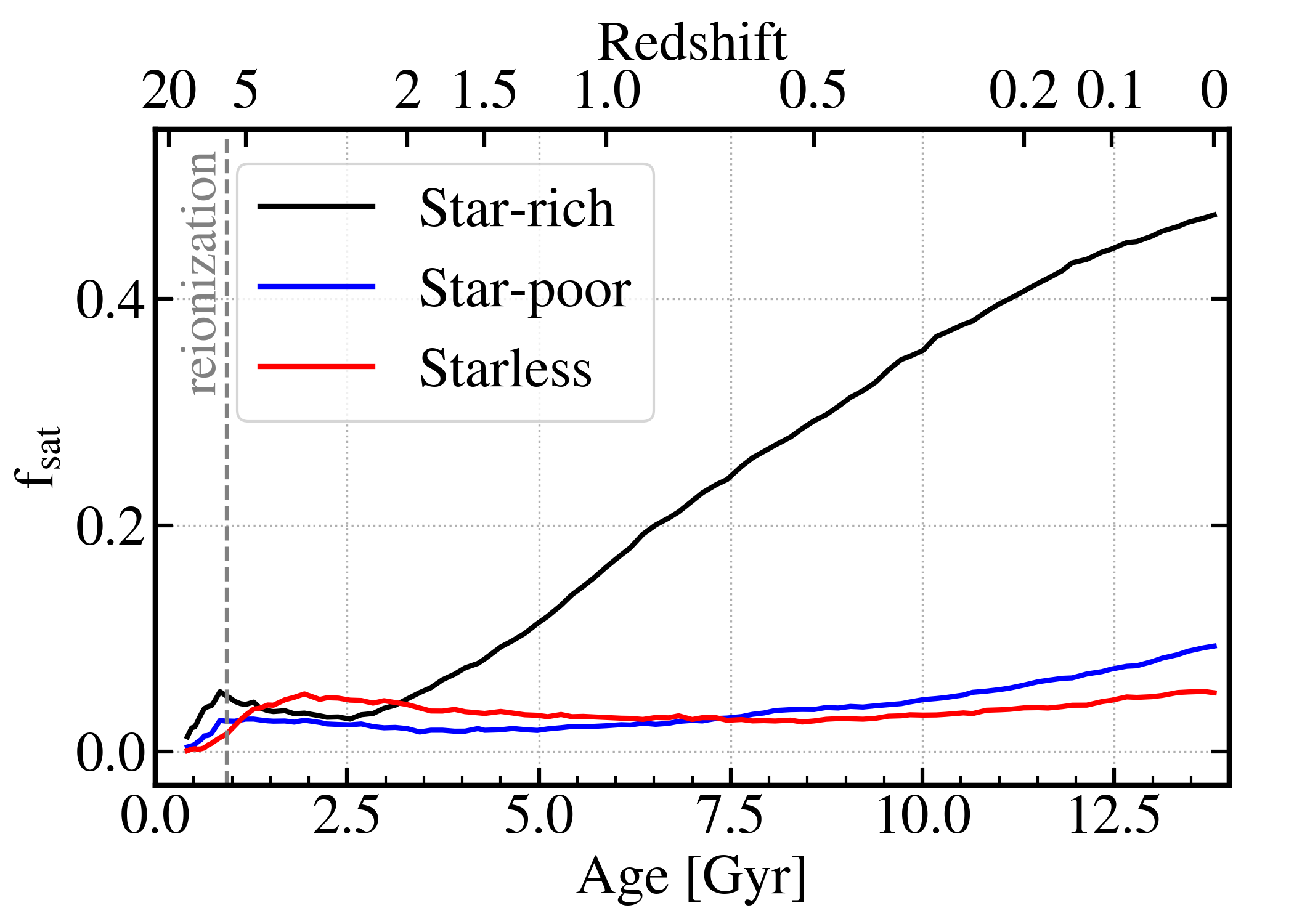}
    \caption{Fraction of satellite galaxies as a function of time for star-rich, star-poor, and starless galaxies. The gray dashed line represents $z=0$, corresponding to the epoch of cosmic reionization.}
    \label{fig:fsat}
\end{figure}

Figure \ref{fig:fsat} shows that the satellite fraction of star-rich galaxies has gradually increased up to approximately 0.5 by $z=0$. This indicates that half of them have become members of galaxy clusters or groups via frequent merging and infalling to the large system. In contrast, the satellite fractions of star-poor and starless galaxies never exceed 0.1 during the entirety of cosmic history, implying that about 90\% of them continue to be isolated. This result provides further evidence supporting the idea that the disparity between luminous and dark galaxies becomes more significant as time goes on.

When galaxies undergo mergers and interactions with other galaxies, they exchange their angular momentum. The process involves the transfer of orbital angular momentum from less massive galaxies to large central galaxies. As a result, the spin parameter of satellites may decrease, contributing to a reduction in their sizes. Moreover, tidal stripping, as galaxies fall into clusters, can remove both gas and dark matter components \citep{Moore1996, Gnedin2003, Aguerri2009, Smith2016, Niemiec2019}. This phenomenon further contributes to a decrease in their sizes. As evidenced in Figure \ref{fig:merger}, such events tend to be more common among luminous galaxies, resulting in the reduction of sizes for their gas and dark matter components; this is illustrated in Figure \ref{fig:evolution}.

\subsection{Cosmic Reionization} \label{subsec:cosmic reionization}
Our results emphasize the important role of cosmic reionization in the evolution of dark galaxies. By heating up gas components and expelling them from galaxies, cosmic reionization can boost the differentiation between luminous and dark galaxies \citep{Thoul1996, Hambrick2011}.

Galaxies that are significantly influenced by cosmic reionization undergo a reduction in their gas reservoirs, which are essential for star formation. Consequently, these galaxies struggle to form new stars and evolve into dark galaxies. On the other hand, galaxies that withstand the impacts of cosmic reionization and maintain their gas supply continue their star formation activities. These galaxies eventually evolve into luminous galaxies (See Figures \ref{fig:gas_farction} \& \ref{fig:evolution}).

One of the factors that can make a galaxy survive from cosmic reionization is its dark matter mass before this epoch. As shown in the dark matter mass evolution history in Figure \ref{fig:evolution}, luminous galaxies clearly contain more dark matter masses than their dark galaxy counterparts just before $z=6$. This abundance of dark matter in luminous galaxies will provide a deeper gravitational potential, which can reduce the gas loss during the cosmic reionization phase.

\begin{figure}
    \centering
    \includegraphics[width=\linewidth]{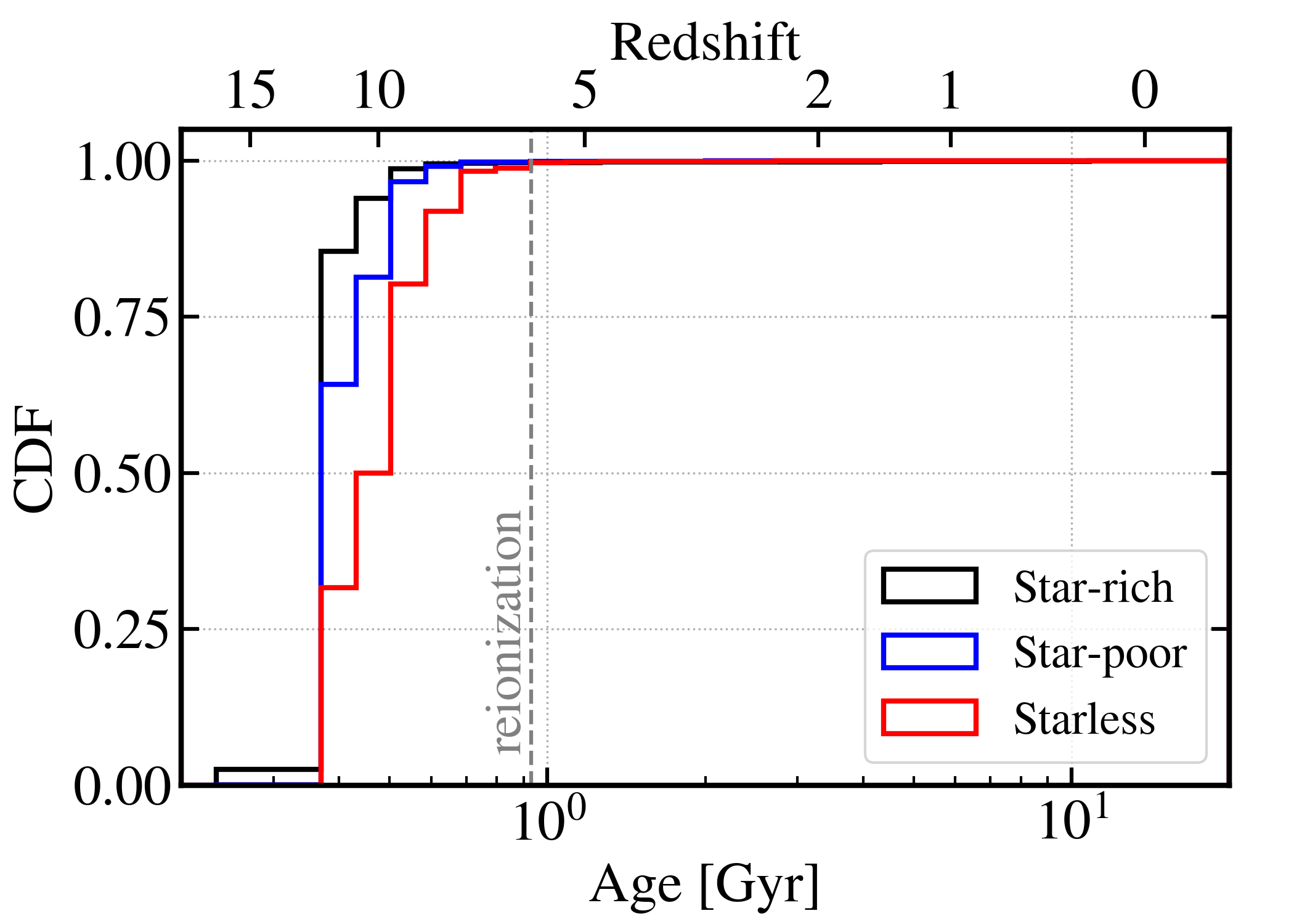}
    \caption{Cumulative distribution function for the time when the galaxies are first identified as SUBFIND groups. The x-axis represents the age of the Universe on a logarithmic scale.}
    \label{fig:form_epoch}
\end{figure}

The formation time of galaxies could be another factor that can make a galaxy survive from cosmic reionization. Figure \ref{fig:form_epoch} shows the time when galaxies are identified as SUBFIND groups; i.e., a group should have at least 20 dark matter and star particles. It reveals that the majority of galaxies are identified before cosmic reionization. However, star-rich galaxies are generally identified earlier than star-poor and starless galaxies. Therefore, dark galaxies could not have enough time to be stabilized, which can make them more vulnerable to the effects of cosmic reionization from outside. As previously mentioned in Section \ref{subsubsec:Mass}, when it comes to formation time, star-rich galaxies require 1.2 Gyr to reach half of their final dark matter mass, while star-poor and starless galaxies need 2.2 Gyr and 3.5 Gyr, respectively. This further supports the idea that luminous galaxies form earlier so that they could be stabilized to survive from cosmic reionization effect.

However, it is important to be aware of the limitations of simulations in modeling cosmic reionization. In the TNG simulation, cosmic reionization starts to operate at $z=6$ based on the assumption of a uniform UV background radiation field as a heating source. Only the gas cells with a density greater than 0.001 $\rm cm^{-3}$ are able to cool via self-shielding. Gas cells with a density lower than this threshold experience heating up to $\rm 10^4\,K$ without cooling from self-shielding. Therefore, in this simple prescription, the effect of ionization could be exaggerated in low-density environments where the density of ionizing photons is also low \citep{Rosdahl2018}. Additionally, it could suppress gas infall onto low-mass galaxies by photoheating the intergalactic gas and raising its pressure \citep{Shapiro1994, Gnedin2000, Hoeft2006}.

\subsection{Other Factors}
There may be other factors that we did not explore in detail, but could still be important for understanding the origin and fate of dark galaxies in the cosmic landscape. We briefly list them here and leave them for future studies.

\textbf{Baryonic feedback:} Baryonic feedback could have an impact on the formation of dark galaxies. For instance, the supernova (SN) feedback mechanism injects energy into gas, affecting both their kinematics and thermodynamics. This energy injection can lead to the suppression of star formation and potentially play a role in the formation of dark galaxies.

\textbf{Ram pressure stripping:} When a galaxy falls into the center of a galaxy cluster, it experiences strong ram pressure due to the surrounding gas \citep{Gunn1972, Hwang2018}. This pressure can remove the gas component of a galaxy, which in turn hinders star formation and may lead to the birth of dark galaxies. According to our findings, roughly 10\% of star-poor galaxies are satellites within galaxy clusters or groups. We suggest that some dark galaxies in relatively dense regions might have formed through this process. These galaxies, which experienced ram pressure as satellites, could have different origins compared to dark galaxies in lower-density environments.

\textbf{Dark matter halo concentration:} Although we have examined the density of dark matter halos in Section \ref{fig:evolution}, it is important to note that this density is calculated within the half-mass radius. To gain further insights into the characteristics of dark matter halos, it is essential to investigate their concentration, as they can vary even when the integrated densities appear similar.

\textbf{Dark matter halo shape:} The shape of the dark matter halo itself could be one of the other factors impacting the formation of dark galaxies. While luminous galaxies seem less perturbed by cosmic reionization than dark galaxies, the disparity might stem from differences in halo configurations. Although both types of galaxies have similar halo masses, dark galaxies might possess more triaxial halo shapes than luminous galaxies, making them more prone to losing their gas.

\section{Conclusion}
Here, we have investigated the nature of dark galaxies using the IllustrisTNG simulation. To do that, we first identify dark galaxies that have a stellar mass less than 0.0001 of their total mass, while luminous galaxies have a stellar mass larger than this criterion. We further subdivide dark galaxies into two groups: star-poor and starless galaxies. Star-poor galaxies have positive stellar masses, while starless galaxies have zero stellar masses. To ensure a fair comparison, we select the galaxies with dark matter mass of $1.0 \times 10^9 \, h^{-1} \text{M}_\odot \leq M_{\text{DM}} \leq \, 3.0 \times 10^9 \,  h^{-1} \text{M}_\odot$. We then compare the physical properties of these populations to understand the formation and evolution of dark galaxies. 

The main findings of our work can be summarized as follows:
\begin{enumerate}
    \item Present-day luminous and dark galaxies have distinct characteristics. Dark galaxies are predominantly found in lower-density regions, often in isolation. They have larger spin parameters than luminous galaxies, and also have larger gas and dark matter sizes than luminous galaxies.

    \item The differences between present-day luminous and dark galaxies originate from their early environments, such as their birthplace. Galaxies formed in less dense regions with a scarcity of star-forming gas tend to become dark galaxies, while those formed in dense regions with sufficient star-forming gas are likely to be luminous galaxies. The initial spin of galaxy halo also seems to correlate in the formation of dark galaxies, with initial dark galaxies exhibiting slightly larger spin parameters than luminous galaxies.

    \item The differences between luminous and dark galaxies become significant as galaxies evolve. One contributing factor is the occurrence of merging events. Luminous galaxies, as located in denser region, experience a greater number of mergers than dark galaxies which reside in isolation. During these mergers and interactions, luminous galaxies can lose their angular momentum, resulting in relatively smaller sizes and higher densities. On the other hand, isolated dark galaxies maintain higher angular momentum throughout their evolution, resulting in relatively larger sizes and lower densities.
    
    \item Cosmic reionization also plays an important role in influencing the destiny of galaxies. It significantly affects the evolutionary history of dark galaxies, particularly impacting their gas properties. This includes processes like gas heating and loss, which prevent dark galaxies from forming stars. On the other hand, luminous galaxies are survivors of these effects probably because of their greater dark matter masses and earlier formation times.
    
\end{enumerate}

Our results suggest that the detection of dark galaxies with X-ray surveys could be challenging due to their low gas temperatures ($T < 10^5$K). Instead, radio surveys like ALFALFA, LITTLETHINGS, and HI survey using FAST could be valuable tools for identifying dark galaxy candidates (M. Kwon et al. in preparation).

Upcoming surveys, such as the Square Kilometre Array (SKA), hold the potential to detect more dark galaxy candidates and expand our understanding of their properties. Additionally, the EUCLID survey, which will provide dark matter lensing maps, can be utilized to identify dark galaxies. Comparing the properties of dark galaxy candidates from these surveys with our results can contribute to validating whether these candidates are indeed pure dark galaxies or not.

In conclusion, our study enhances our understanding of dark galaxies and highlights their unique characteristics and evolutionary history.


\begin{acknowledgments}
We thank the referee for the constructive comments that improve the manuscript.
We thank the IllustrisTNG collaboration for making their simulation data publicly available.
HSH and HS acknowledge the support by the National Research Foundation of Korea (NRF) grant funded by the Korea government (MSIT), NRF-2021R1A2C1094577 and NRF-2022R1A4A3031306, respectively.
We thank Chang-Goo Kim, Juhan Kim and Changbom Park for useful discussion.
\end{acknowledgments}

%

\vspace{5mm}




\bibliography{refs}
\bibliographystyle{aasjournal}



\end{document}